# Vector Control Algorithm Based on Different Current Control Switching Techniques for Ac Motor Drives

*Muhammad Usama*[1,*], and *Jaehong Kim*[1]

[1]Electrical Engineering Department, Chosun University, 61452 South Korea

**Abstract.** A comparative analysis of vector control scheme based on different current control switching pulses (HC, SPWM, DPWM and SVPWM) for the speed response of motor drive is analysed in this paper. The control system using different switching techniques, are comparatively simulated and analysed. Ac motor drives are progressively used in high-performance application industries due to small size, efficient performance, robust to torque response and high power to size ratio. A mathematical model of ac motor drives is presented in order to explain the numerical theory of motor drives. The vector control technique is utilized for efficient speed control of ac motor drive based on independent torque and air gap flux control. The study compares the total harmonic distortion contents of phase currents of ac motor drive and speed response in each case. The simulation result shows that total harmonic distortion across the phase current in SVPWM is less as compared to other switching techniques while the rise time in speed response across SVPWM technique is faster as compared to other switching methods. The simulation result of ac motor drives speed control is demonstrated in Matlab/Simulink 2018b.

## 1 Introduction

Industrial application required efficient and accurate performance from motor drives. In past dc machines are mostly used in industrial application as controlling dc machine are quite easy. Before the 1950s all industrial appliances use the dc motor drives. After the development of vector control method for ac motor drive, in 1968 open the gateway for future research and efficient control of the motor drive. Ac motor drives are further classified based on external dc current. (1) Induction Motor (2) Permanent Magnet Synchronous Motor Asynchronous machines are widely used in industry but in recent years synchronous machines catch the attention and applied in several different areas of industry such as Electric vehicles, robotics, aerospace technology, and traction respectively. PMSM has several advantages such as small size, high torque to inertia ratio, high power to weight ratio and high efficiency [1].

There are two control techniques for controlling the ac motor drive (1) scalar control (2) vector control [2]. Scalar control just depends upon the magnitude and doesn't contain any phase information of motor drive so because of that it is not valuable for variable speed application. Whereas the vector control method is a closed-loop feedback control method, based on both magnitude and angle of each phase current and voltage control that provide an efficient result at a standstill and transient response.

In the vector control method [3], the motor torque and flux are controlled separately based on PI controllers that make the ac motor drive control in the same sense as controlling the dc machine. The state variables or the control variable that must be controlled are transformed from the stationary reference frame to an exciting reference frame that makes the control simple and efficient.

With the advancement in semiconductor technology, the speed control of ac motor drive becomes easy. Inverter with semiconductor switches (MOSFET, IGBT, etc.) creates variable frequency from the direct current source which is further utilized to drive ac motor drive. The two types of inverter exist with different structure and totally different behaviour. Voltage source inverter and current source inverter. Both these inverter type are totally different from each other based on there working principle [4], [5]. VSI operates on current control mode fed by stiff dc voltage whereas the CSI operates on voltage control mode fed by the stiff current source.

Different switching technique has been utilized for controlling the on-off mode of inverter and switching time of semiconductor switches in the inverter[6]. Hysteresis controller, sinusoidal pulse width modulation, Discontinuous pulse width modulation and space vector pulse width modulation techniques are briefly discussed in this paper for different speed response of ac motor drives & total harmonic distortion cross the phase current.

## 2 Ac Motor Drives

### 2.1 Induction motor

Asynchronous machines are also known as Induction motor. Induction motors are easy to construct and

---

[*] Corresponding author: jaehong77@gmail.com

numerous advantages such as less maintenance required, low cost, compactness, ease of use and robustness [7]. Most asynchronous machines are rotary type with rotary rotor and stationary stator. When the stator winding of induction motor is driven by three-phase supply a magnetic field with constant magnitude and rotating at electrical speed is produced, causing the electromotive force (EMF) to induce in rotor conductor. The rotor conductor ends are short-circuited on themselves and the induced electromotive force sets up the current in rotor conductor in such a direction as to produce a torque. Which rotates the rotor in the same direction as the magnetic field. The rotor speed tends to increase gradually and tries to reach the speed of the exciting magnetic field, but it can't reach the desired speed otherwise no EMF will be induced across rotor conductor and torque will become zero. So rotor rotates with rotor speed that is very close to synchronous speed but not equal. The equivalent circuit of induction motor is shown in *Fig. 1*

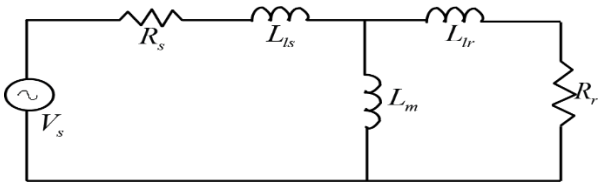

**Fig. 1** Per-Phase Equivalent Circuit of IM

The mathematical model of stator and rotor voltage equation are given in (1) and (2) with the following assumption that mutual inductances are equal while the eddy current, hysteresis loss and saturation effect is ignored

$$Vs = (R + j\omega_e L_{ls})I + (j\omega_e L_m)(I + I_r) \quad (1)$$

$$0 = \left(\frac{R_r}{s} + j\omega_e L_{lr}\right)I_r + (j\omega_e L_m)(I + I_r) \quad (2)$$

Where $R$ and $R_r$ the stator and rotor resistance, $\omega_e$ is the electrical speed, $L_{ls}$ and $L_{lr}$ are the stator and rotor leakage inductance and $L_m$ is the magnetizing inductance.

$$T_e = \frac{P_{out}}{\omega_e} = \frac{[3I^2 R_r[1-S]]}{\omega_e S} \quad (3)$$

$$S = \frac{\left[\omega_e - \frac{p}{2}\omega_r\right]}{\omega_e} \quad (4)$$

Where S is the slip speed and is given in (3). The output power is the multiplication of electromagnetic torque and electric speed so is given as

$$P_{out} = \frac{[3I^2 R_r[1-S]]}{s} \quad (5)$$

**2.2 Permanent Magnet Synchronous Motor**

Industrial applications require efficient and accurate performance from motor drives. Asynchronous machines are widely used in industry but in recent years synchronous machines catch the attention and applied in several different areas of industry such as Electric vehicles, robotics, aerospace technology, and traction respectively. PMSM has several advantages such as small size, high torque to inertia ratio, high power to weight ratio and high efficiency.

In the three-phase coordinate system, the voltage equation of three-phase SM-PMSM is derived as [1].

$$\begin{bmatrix} v_a \\ v_b \\ v_c \end{bmatrix} = r_s \begin{bmatrix} i_a \\ i_b \\ i_c \end{bmatrix} + \frac{d}{dt} \begin{bmatrix} \lambda_a \\ \lambda_b \\ \lambda_c \end{bmatrix} \quad (6)$$

where $v_{abc}$, $i_{abc}$, $\lambda_{abc}$, $r_s$ denote the motor phase voltage, phase current, flux linkage in stator windings, and stator winding resistance, respectively. With the Park's transformation (6) is transformed to the synchronous reference frame as

$$\begin{bmatrix} v_d^e \\ v_q^e \end{bmatrix} = \begin{bmatrix} r_s L \, \mathcal{P} & -\omega_e L \\ \omega_e L & r_s L \, \mathcal{P} \end{bmatrix} \begin{bmatrix} i_d^e \\ i_q^e \end{bmatrix} + \begin{bmatrix} 0 \\ \omega_e \lambda_m \end{bmatrix} \quad (7)$$

where $\omega_e$ and $\lambda_m$ are the electrical angular velocity and magnetic flux linkage caused by the rotor side permanent magnet, respectively $L_s$ is the stator inductance, and $\mathcal{P}$ is the differential operator respectively. The torque equation of PMSM is

$$T_e = \frac{3P}{2}[\lambda_d i_q - \lambda_q i_d] \quad (8)$$

where $T_e$ and P are generated electrical torque and the number of motor poles, respectively. The mechanical equation of PMSM is

$$T_m = T_L + B\omega_m + J\mathcal{P}\omega_m \quad (9)$$

where $T_m, T_L, B, J$ and $\omega_m$ are mechanical torque, load torque, viscous friction, inertia and mechanical angular frequency. The DQ-axis dynamic model of PMSM is derived to evaluate the vector control of PMSM. Stator voltage equation under synchronous DQ-reference frame is written in (7) by neglecting the eddy current and hysteresis losses as well as there no cage on the rotor and induced emf is sinusoidal [8].

**2.3 Vector Control**

Vector control also called field-oriented control is an adjustable speed drive control method in which a three-phase stator current is transformed into two-phase identified as two orthogonal components that can be defined as a vector. In vector control method both the torque and flux are decoupled from each other so it provides an easy way to control both torque and flux independently [9]. The mathematical model of vector control is complex, but it provides precise control of ac motors as well as excellent dynamic response and high performance with less loss of power.

## 2.4 DQ Transformation

The Clark and Park transformations are done for the analysis of motor parameters at the stationary and rotating frame. Clark transformation is a technique to convert three-phase supply to the alpha-beta stationary reference frame, whereas the park transforms converts three phases to the direct-quadrate(dq) rotating reference frame. The dq transformation is shown in *Fig. 2*. The mathematical model for dq transformation is given as

$$\begin{bmatrix} v_\alpha \\ v_\beta \\ v_0 \end{bmatrix} = K \begin{bmatrix} 1 & -\frac{1}{2} & -\frac{1}{2} \\ 0 & \frac{\sqrt{3}}{2} & -\frac{\sqrt{3}}{2} \\ \frac{1}{\sqrt{2}} & \frac{1}{\sqrt{2}} & \frac{1}{\sqrt{2}} \end{bmatrix} \begin{bmatrix} v_a \\ v_b \\ v_c \end{bmatrix} \quad (10)$$

Where $K = \frac{2}{3}$

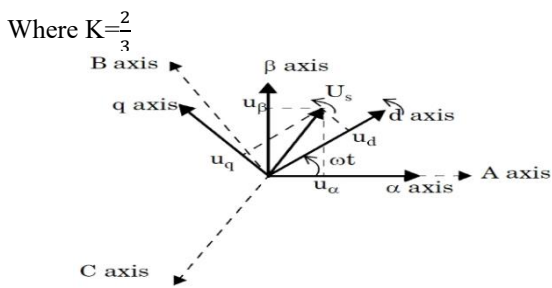

**Fig. 2** DQ Transformation

Park Transformation is given as:

$$\begin{bmatrix} v_d \\ v_q \\ v_0 \end{bmatrix} = K \begin{bmatrix} \cos\theta & \cos\left(\theta - \frac{2\pi}{3}\right) & \cos\left(\theta - \frac{2\pi}{3}\right) \\ \sin\theta & \sin\left(\theta - \frac{2\pi}{3}\right) & \sin\left(\theta - \frac{2\pi}{3}\right) \\ \frac{1}{\sqrt{2}} & \frac{1}{\sqrt{2}} & \frac{1}{\sqrt{2}} \end{bmatrix} \begin{bmatrix} v_a \\ v_b \\ v_c \end{bmatrix} \quad (11)$$

## 2.5 PI Controller

Speed control of ac motor drive is governed by two sets of the controller, the inner controller that is the current controller (ACR) that is used to control the current based on voltage whereas the outer controller is the speed controller (ASR) utilized to control the torque. In field-oriented control encoder is used to provide the rotor speed and position information in closed-loop whereas in direct torque control method (DTC) the position signal is not required for speed control but it causes high torque ripple, so because of that mostly field-oriented control technique is utilized as it is efficient, reliable and provide good speed response with less torque ripple etc. is implemented in industries application. The mathematical equation for the PI controller is given as

$$y = K_p + \frac{K_i}{s} \quad (12)$$

## 3 Inverter Switching Techniques

### 3.1 Hysteresis Current Controller

In field-oriented control, torque depends upon stator current vector and this shows that the accurate speed control be determined, by how well stator current vector is controlled. The vector control block diagram of hysteresis band controller is shown in *Fig. 3*. In high-performance motor drives, the hysteresis controller exempts the use of current regulator and compare the real stator current with the reference current value and the error signal is fed to comparator having hysteresis band. When the error signal crosses the upper limit of hysteresis band the lower switch turns one whereas when the error signal crosses the lower limit of hysteresis band the upper leg of inverter switch turns on [10]. In other words, if the measure phase current is greater than or less then reference current then the pulse sequences are generated to drive the MOSFET switches.

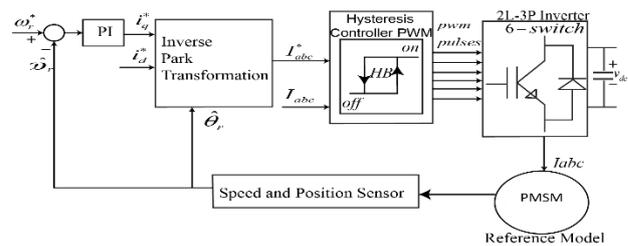

**Fig. 3** Hysteresis Controller Vector Control Block Diagram

a) If $I_{ref} - I_{act} \leq$ -H the lower switch will be turned on caused the load current to decrease by producing negative load voltage $-\frac{1}{2}V_{dc}$

b) If $I_{ref} - I_{act} \geq$ H then the upper switch will be turned on cause the load current to increase by producing the positive load voltage $\frac{1}{2}V_{dc}$

In the hysteresis controller the stator phase voltage switches to maintain the stator phase current within hysteresis band. The smaller the hysteresis band the more closely stator phase current represents sinusoidal wave, but small hysteresis band causes high switching frequency which leads to inverter loss. In hysteresis controller the motor parameter, the comparator and the operation conditions depend upon the switching frequency which results in variable switching frequency and that's the major drawback of this control scheme because in wide speed range the variable frequency is not effective.

### 3.2 Sinusoidal Pulse Width Modulation

Pulse Width Modulation is a DC voltage modulation method that provides intermediate voltage by adjusting the on and off ratio in each period, in inverter application the alternating current output voltage and direct current input voltage varies continuously, so to achieve desired output voltage waveform the continuous variation of control pulse width is required. SPWM based vector control system block diagram is shown in *Fig. 4* .Sine pulse width modulation is one of the renowned methods for controlling the inverter output voltage. SPWM is a simple and quite easy method to be implemented. As compared with the hysteresis control system, SPWM vector control system is double-loop structure. The SPWM is realized by comparing the sine wave reference signal with high frequency triangular carrier signal that results in the inverter switching time instant, which generates the pulse sequence to drive the switching devices [11].

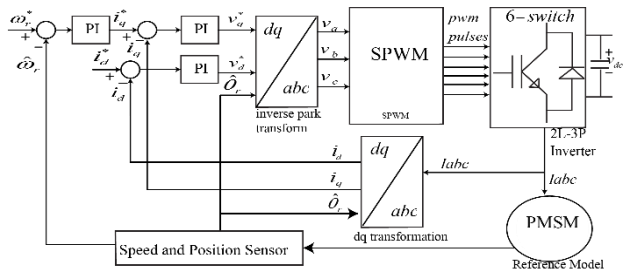

**Fig. 4** SPWM Vector Control Block Diagram

*Fig. 5* shows the waveform of reference signal and carrier signal along with sequence of generated switching pulse The harmonic content in the inverter voltage output waveform is determined by the modulation index that is the ratio of the reference voltage to the carrier signal [12].Moreover, the frequency of reference signal determines the inverter output frequency and its peak amplitude determines the modulation index which in turn determines the RMS output voltage. The total harmonic distortion is relatively high in SPWM. Moreover, to increase the peak value of RMS voltage some high order harmonics are added. In other words, the over-modulation method is to be performed to increase the peak RMS voltage value.

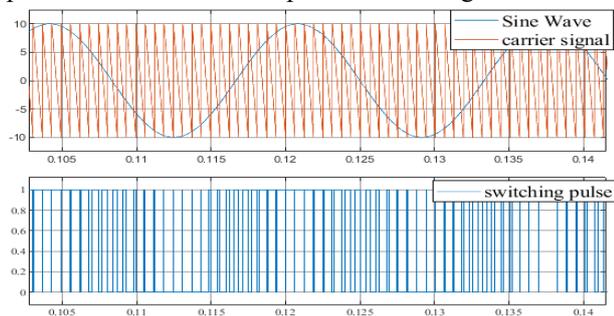

**Fig. 5** Pulse Order of SPWM

### 3.3 Discontinuous Pulse Width Modulation

DPWM uses the discontinuous type of zero sequence signal. In discontinuous PWM technique for each time sample, one phase ceases the modulation and the associated phase is clamped to negative or positive dc bus. The PWM performance depend upon modulation index. In lower modulation range, the CPWM techniques are better than DPWM technique while in higher modulation range DPWM is superior[13][14]. The block diagram of DPWM mechanism is shown in *Fig. 6*. and the DPWM mechanism block is replaced with SPWM in Simulink model for switching control.

Where k is given as k=1-$\alpha$

$$\alpha = \frac{1}{2}[1 + sgn(cos3(\omega t + \delta + \varphi)]$$

For unity power factor $\varphi$= 0,

Variation of Modulation phase angle $\delta$ yield to infinite DPWM techniques.

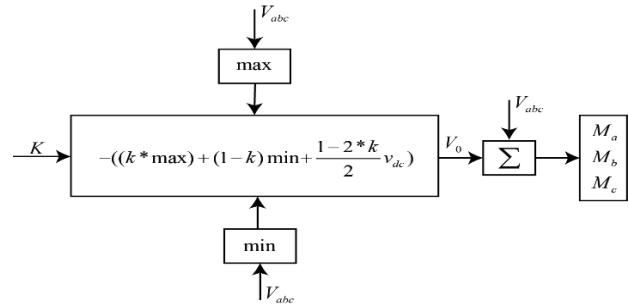

**Fig. 6** DPWM Mechanism

### 3.4 Space Vector Pulse Width Modulation

The SVPWM VSI is used as it offers a 15% increase in DC-link voltage utilization and low order harmonic distortion as compared to standard sinusoidal PWM VSI. Space vector PWM reflects a special switching sequence for the three-phase inverter. For implementing the space vector PWM the reference voltages are transformed into stationary alpha-beta reference frame by the matrix transformation based on Clark transformation as shown in (13). Based on given eight switching pulse patterns where six are the non-zero voltage vector and two are zero voltage vector [15].The amplitude of the non-zero switching pulse is $\frac{2}{3}V_{dc}$ and the phase difference of 60 degrees. The non-zero voltage vectors are placed on the edge of hexagon and two zero voltage vector are placed on the origin as shown in *Fig. 7*.At V0 and V7, no dc-link voltage is applied across the load.

$$\begin{bmatrix} v_\alpha \\ v_\beta \end{bmatrix} = \begin{bmatrix} \cos\theta & -\sin\theta \\ \sin\theta & \cos\theta \end{bmatrix} \begin{bmatrix} v_d \\ v_q \end{bmatrix} \quad (13)$$

Space Vector PWM can be implemented by the following steps:

a) determine the stationary Frame reference voltages

b) Find the time duration T1, T2, T0 that are given as

c) Find the switching Time of each Switch of the power inverter.

Switching time calculation at each Sectors is calculated based on (14). Based on (14) SVPWM is modeled in Matlab and output is shown in *Fig. 8*

$$T_1 = \frac{\sqrt{3}T}{V_{dc}}[\sin\frac{\pi n}{3}v_\alpha - \cos\frac{\pi n}{3}v_\beta] \quad (14)$$

$$T_2 = \frac{\sqrt{3}T}{V_{dc}}[\sin\frac{\pi(n-1)}{3}v_\alpha - \cos\frac{\pi(n-1)}{3}v_\beta]$$

$$T_0 = T_s - [T_1 + T_2]$$

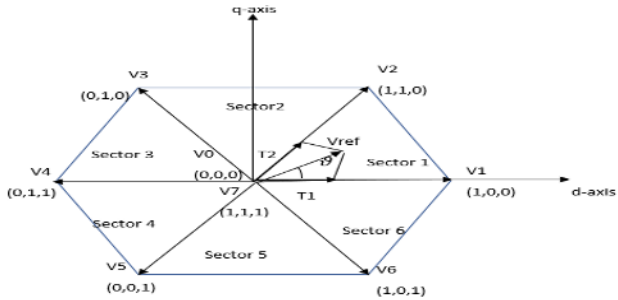

**Fig. 7** Basic Switching Vector and Sectors

In SVPWM the total harmonic distortion is reduced and no need for over-modulation as the SVPWM switching signal contains the high order harmonics as shown *Fig. 8*. So RMS voltage value is higher than sinusoidal pulse width modulation. Moreover, the space Vector pulse width modulation provide more efficient and reliable use of phase supply voltage as compared to sinusoidal pulse width modulation

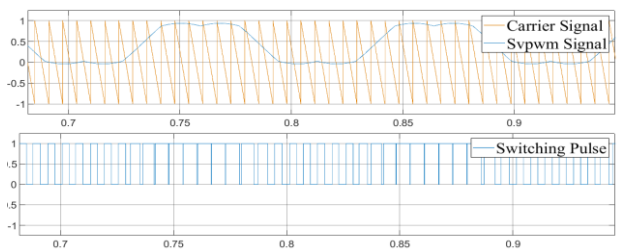

**Fig. 8** Pulse Order of SVPWM

## 4 Results and Discussion

In MATLAB/Simulink workspace, the vector control techniques based on different switching pulses algorithm were built. The speed and the current regulator were driven by PI controllers. Synchronous machines are widely used both in industry and home applications due to small size, high power density and efficient performance, so in this paper SPMSM is utilized for modeling and simulation results. Motor parameters are given in **Table 1** that in SVPWM switching technique the rise time of speed response is faster as compared to other switching techniques. The overall performance of the system is fast and good. FFT analysis on the phase current of the speed control model with fundamental frequency equaling 50Hz is analyzed and shown in *Fig. [10-13]*. The THD across phase current under different load conditions are shown in **Table 2.** So the less THD in SVPWM across the load makes it an efficient technique for closed-loop speed control of motor drive under variable load conditions.

**Table 1.** Simulation Parameter

| Parameters | Value |
|---|---|
| $T_s/T_{pwm}$ | 1us/100us |
| Stator Resistance | 0.675 ohm |
| Stator Inductance | 0.000835 |
| Pole pairs | 4 |
| Inertia | 0.01 |
| Rotor Flux Linkage | 0.11 |

*Fig. 9* Shows the speed response during startup at 0s under 5Nm load condition. At 0.3s the speed is increased from 100rad/s to 300rad/s. At 1s the load torque of 8Nm is applied. Simulation result shows

**Table 2**. THD Across Phase Current

| $T_L$ | Time(THD) | HPWM | DPWM | SPWM | SVPWM |
|---|---|---|---|---|---|
| 5Nm | 0.8s | 29.57% | 13.05% | 13.03% | 9.75% |
| 8Nm | 1.9s | 16.97% | 9.87% | 10.14% | 8.08% |

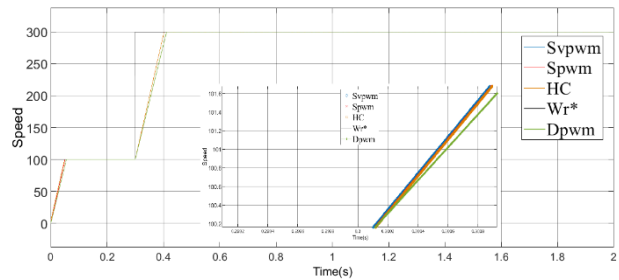

**Fig. 9** Speed Response

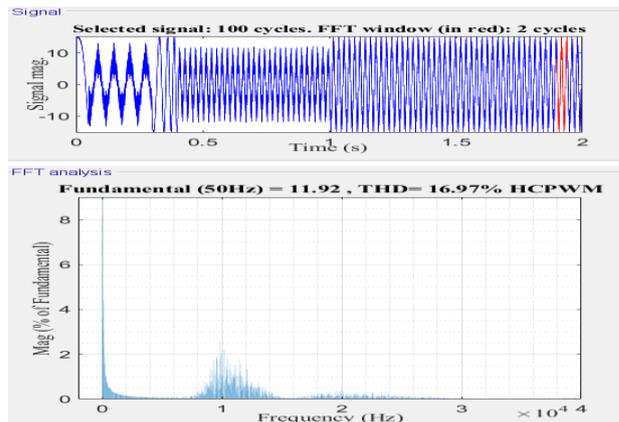

**Fig. 10** FFT analysis of HCPWM

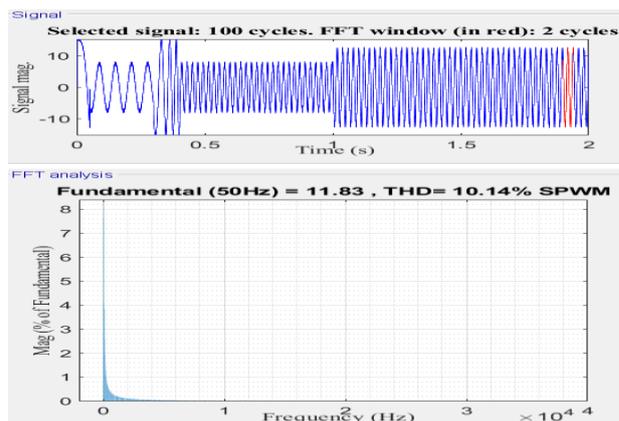

**Fig. 11** FFT analysis of SPWM

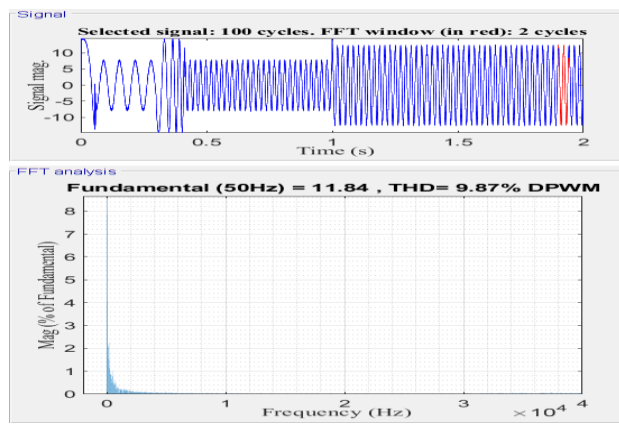

**Fig. 12** FFT analysis of DPWM

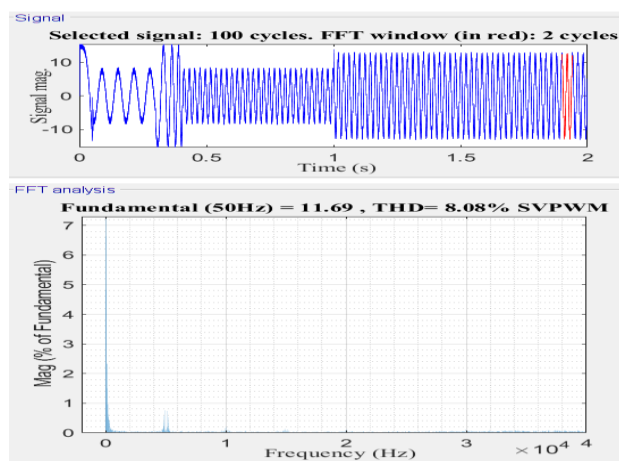

**Fig. 13** FFT analysis of SVPWM

## 5 Conclusion

Vector control of ac motor drive based on different current control switching techniques are implemented in Matlab/Simulink 2018b. To study the dynamic response of ac motor drive both transient and standstill conditions are defined and model in software. The simulation result shows that the SVPWM algorithm has lower THD and fast rise time for speed response while the SPWM, HC and DPWM has slow rise time for speed response as well as higher THD. This study helps to find the better and efficient switching technique for ac motor drive speed control and utilized in sensorless control for future work. We conclude that Matlab/Simulink is an efficient and reliable tool to analyse and predict the response of ac motor drive based on different switching techniques.

This work was supported by the National Research Foundation of Korea (NRF), grant funded by the Korea government (MSIT). NRF-2017R1C1B5018301)